\def\be{\begin{equation}}
\def\ee{\end{equation}}

\def\nc2{\left( \begin{array}{c} n \\ 2 \end{array}\right)}

\def\beq{\begin{equation}}
\def\eeq{\end{equation}}

\def\d{{\cal D}}

\def\n{\noindent}

  \def\ket{\vert \vert  \{ \emptyset \} \rangle}
  \def\ket2{\vert \vert \otimes \{ R \} \rangle}
  
\def\dpr{\prime\prime}

\def\.#1{\mathaccent 95#1}
\def\^#1{\mathaccent 94 #1}
\def\~#1{\mathaccent "7E #1}

\def\equal{\enskip =\enskip}
\def\plus{\enskip +\enskip}

\def\ul#1{\underline{#1}}

  \def\ket{\vert \vert  \{ \emptyset \} \rangle}
  \def\ket2{\vert \vert \otimes \{ R \} \rangle}

\def\d{{\cal D}}

  \def\ket{\vert \vert  \{ \emptyset \} \rangle}
  \def\ket2{\vert \vert \otimes \{ R \} \rangle}

  \def\ket{\vert \vert  \{ \emptyset \} \rangle}
  \def\ket2{\vert \vert \otimes \{ R \} \rangle}
  
\def\dpr{\prime\prime}

\def\.#1{\mathaccent 95#1}
\def\^#1{\mathaccent 94 #1}
\def\~#1{\mathaccent "7E #1}

\def\equal{\enskip =\enskip}
\def\plus{\enskip +\enskip}

\def\ul#1{\underline{#1}}

\newif\ifletter

\newcommand{\boldarrayrulewidth}{1\p@} 
\def\bhline{\noalign{\ifnum0=`}\fi\hrule \@height 
\boldarrayrulewidth \futurelet \@tempa\@xhline}
\def\@xhline{\ifx\@tempa\hline\vskip \doublerulesep\fi
      \ifnum0=`{\fi}}

\newcommand{\ms}{\noalign{\vspace{3\p@ plus2\p@ minus1\p@}}}
\newcommand{\bs}{\noalign{\vspace{6\p@ plus2\p@ minus2\p@}}}
\newcommand{\ns}{\noalign{\vspace{-3\p@ plus-1\p@ minus-1\p@}}}
\newcommand{\es}{\noalign{\vspace{6\p@ plus2\p@ minus2\p@}}\displaystyle}
\newcommand{\etal}{{\it et al\/}\ }

\newcommand{\IJMP}{{\sl Int. J. Mod. Phys.\ }}
\newcommand{\MPL}{{\sl Mod. Phys. Lett.\ }}

\newcommand{\JPB}{{\sl J. Phys. B: At. Mol. Phys.} }      
\newcommand{\JPCM}{{\sl J. Phys.: Condens. Matter\/} }    
\newcommand{\CPL}{{\sl Chem. Phys. Lett.} }

\newcommand{\JPCm}{{\sl J. Phys. Chem.} }   

\newcommand{\SuSc}{{\sl Surf. Sci.\ } }

\newcommand{\JCP}{{\sl J. Chem. Phys.} }

\newcommand{\JPhCh}{{\sl J. Phys. Chem.\ } }

\newcommand{\JPSJ}{{\sl J. Phys. Soc. Japan\/} }

\newcommand{\PR}{{\sl Phys. Rev.} }
\newcommand{\PRL}{{\sl Phys. Rev. Lett.} }

\newcommand{\AS}{{\sl Appl. Spectroscopy}}

\documentclass[12pt,epsfig]{article}
\begin{document}
\begin{center}

{\Large\bf A FIRST PRINCIPLES STUDY OF CU CLUSTERS} 
\vskip 0.5cm
 {\bf Aninda Jiban Bhattacharyya, Abhijit Mookerjee}  \\
S.N. Bose National Centre for Basic Sciences,\\
 JD Block, Sector 3, Salt Lake City,\\ Calcutta 700091,
India\\
{\bf and}\\
{\bf A.K. Bhattacharyya}\\
Department of Engineering,\\ University of Warwick,
Coventry CV47AL, U.K.
\end{center}
\abstract{ \small In this communication we study the equilibrium shapes and energetics
of Cu clusters of various sizes upto 20 atoms using the Full-Potential Tight
Binding Muffin-tin Orbitals Molecular Dynamics. We compare our results with
earlier works by physicists and chemists using different methodologies.}

\section{Introduction}
\normalsize
Many of the first principles molecular dynamics approaches to the study of clusters 
 depend upon the construction of suitable pseudopotentials for the
constituent atoms. Transition metal clusters require perhaps, alternative treatments
 (\cite{kn:lcao2,kn:datni,kn:jena}).
The deep potentials associated with their $d$-orbitals are not particularly amenable
to the pseudopotential approach. In this communication we shall describe a study of Cu clusters
using a full-potential LMTO based molecular dynamics.

Experimentation on the electronic and cohesive properties
of transition metal clusters have been extensive \cite{bb,cc,k,d,gl,h,h2,hm,mh,oz,oz2,sm}. 
The smaller Cu clusters have been exhaustively studied by quantum chemists
\cite{ats,ah,a,a2,bdv,b,bm,bs,bnm,b2,drsv,kt,l,m76,m77,n,p,pb,tmn,ug}.
An excellent early review of the field has
been made by Ozin \cite{ozin}. The main issues addressed in these works were : 
 whether small clusters  had characteristics of bulk metals and in
what way they differed from them in respect to cohesive energies, ionization potentials
and magnetism. Recently Apai \etal \cite{kn:apai} conducted EXAFS studies of Cu clusters
supported on carbon. Similar studies of Au and Ag clusters were carried
out by Balerna \etal \cite{kn:ba} and  Montano \etal \cite{kn:mon}.
These studies indicate, as one would expect for these metals, that the
localized {\sl d-} electrons play an important role in the electronic
structure. Hence, the {\sl d-}states and their interactions with the extended {\sl s-}%
states need to be carefully accounted for in a proper theoretical
treatment of these materials.

We may classify the theoretical approaches into five groups :

\begin{description}
\item[(i)]In the first group are the Hartree-Fock and $X\alpha$ descriptions
(\cite{pb}).  In this class we have the the self-consistent-field-$X\alpha$ (\cite{m77}),
the ab-initio self-consistent-field (SCF) using model
potentials (\cite{bnm}) and the SCF with relaxation effects (\cite{miyo,tmn}).

\item[(ii)] In the next group are the methods based on the local
spin density (LSDA)  both without and including self-interaction corrections (\cite{wang}).
Salahub and coworkers have argued that it is essential to include the gradient
corrections in the LSDA in order to treat clusters properly 
(\cite{cs,fz,fas,gs,ss,sscd}) since the bonding charge density in a small cluster is highly inhomogeneous.

\item [(iii)] The third group includes tight-binding type methods. 
These include the extended H\"uckel methods (\cite{a3,bm,b78}), re-parameterized H\"uckel with
the Wolfsberg-Helmholtz approximation for the off-diagonal terms (\cite{m76}) and those 
with more flexible forms for them (\cite{amos,j82}). This group also has the linear combination
of atomic orbitals based SCF methods (\cite{cv}). We also have the  empirical tight-binding (TB) or
the Linear Combination of Atomic Orbitals (LCAO)
methods (\cite{kn:lcao}, \cite{kn:lcao2}). These are at best qualitative,
since the assumption of transferability of the Hamiltonian
parameters is definitely of questionable validity.

\item [(iv)] In the fourth group we have 
the effective potential methods which include the embedded atom pair and many-body
potentials (\cite{cl91,rfft}) and the effective medium theory (\cite{h89,j87}) with
one-electron correlation included (\cite{ckrs,k91,v92}). The equivalent crystal theory
(ECT) (\cite{kn:smith,kn:ban,kn:perry}) also belongs to this class of empirical potential
methods and is capable of dealing with very large clusters.

\item[(v)] Finally we have attempts at using the tight-binding 
linearized muffin-tin orbitals (TB-LMTO)                
method (\cite{kn:lmto}) coupled to simulated annealing. In the
application of this method to clusters there are several                 
outstanding problems. The treatment of the interstitial region                  
outside the muffin-tin spheres centered at the atomic positions is       
difficult. Unlike the bulk, where the interstitial region is 
small and inflating the muffin-tins to slightly overlapping
atomic spheres can do away with the interstitial altogether, for                  
clusters this is certainly not so. As the atoms move about, the
atomic spheres may not overlap and the interstitial contribution                                    
is significant. One may try to overcome this by enclosing the
cluster with layers of empty spheres carrying charge but not
atoms. This complicates the actual calculations and the justification
of extrapolation of the TB-LMTO parameters beyond the {$5\%$}
range on either side of the equilibrium value is not valid.

\end{description}

 A number of molecular dynamics studies of the geometrical and electronic
structure of small clusters of various elements (\cite{kn:mars, kn:kon,kn:vijay}) have been performed. 
The {\sl ab-initio} molecular dynamics (MD)
approach developed by Car and Parinello \cite{kn:cp} (CP) has been one of
the most promising developments in this area. 
 The method is based on
the pseudopotential technique and therefore faces problems when dealing with
the rather localized \d electrons of transition metals. Efficient soft  
pseudopotentials for transition metals are still not available and the
CP generally is never applied to transition metal clusters.

Simple alkali metals clusters are fairly well described by the
spherical jellium model. The quasi-free valence electrons occupy
single-particle states in an effective spherically symmetric box
potential. This is rather insensitive to the geometry of the
atomic arrangement inside the cluster. Consequently one obtains a
pronounced shell closing effect (\cite{cl91}). Although the noble metals Cu, Ag
and Au have closed d-shells and singly occupied outermost s-shell
structures and several authors have suggested that there should
be a close similarity to the shell closing effect in simple
alkali metals, cohesive studies in the bulk metal and a series of
EXAFS studies of Cu clusters supported on
carbon (\cite{kn:apai},\ \cite{kn:mon})
indicate that the {\sl d-}electrons through their hybridization with
the {\sl s-}electrons play an important role in the electronic
structure and binding energy of these systems. \cite{wpr} have also indicated
through a series of experiments which include mass spectroscopy, oxygen and
water absorption, that there is a competition of jellium-like electronic
behaviour and icosahedral  geometrical closure effects in small copper
clusters.

In this chapter, we shall turn to the molecular version of
the full-potential linearized muffin-tin orbital two-centre-fit (TCF) method
 suggested by 
 \cite{kn:ms} and \cite{kn:mss}
to carry out an {\sl ab initio}     
study of Cu clusters ranging in size between 10 and 20.

\section{The two-center fit method : TCF}

The molecular version of the full potential-LMTO two-centre-fit (TCF)
method  utilizes the
philosophy of Muffin-Tin Orbitals methods. It is based on
the Density Functional Theory in the Local Density Approximation.
The electron-electron interaction is treated approximately. In
practice :

\begin{equation}
\left[ -\nabla^{2}\plus V_{eff}(\ul{r})\right] \psi_{i}(\ul{r}) 
\equal  \varepsilon_{i} \psi_{i}(\ul{r}) \end{equation}

\n where,

\[ \rho(\ul{r})  \equal  \sum_{i} f_{i} \vert
\psi_{i}(\ul{r})\vert^{2} \]

\n and,

\[ V_{eff}  \equal  V_{N}(\ul{r})+2\int
\frac{\rho(\ul{r'})}{\vert r-r' \vert } d^{3}\ul{r'} +
\mu_{xc}\left(\rho(\ul{r})\right)
\]

The first step is the solution of the Schr\"odinger equation in a
very {\sl unpleasant} potential with Coulomb singularities. As in
most approaches we use the variational approach. We choose a
basis of representation $\{ \phi_{m}(\ul{r}) \}$ such that

\[ \Psi(\ul{r}) \equal \sum_{m} c_{m} \phi_{m}(\ul{r}) \]

The problem reduces to a matrix eigenvalue problem :

\[ {\cal{H}} \ul{c} \equal \varepsilon {\cal{S}} \ul{c} \]

Computation effort scales as $\sim$ (matrix dimension)$^{3}$. Our
approach tries to use a minimal basis set at the expense of a
rather complicated formulation. The basis is built up of H\"ankel
functions H$_{iL}$ diverging at $\ul{r}$ = $\ul{R}_{i}$,
augmented inside the muffin-tin spheres by  solutions
u(r)Y$_{L}(\^r)$ of the Schr\"odinger equation :

\[ u_{L}^{\dpr}(r)\equal \left[\frac{\ell(\ell+1)}{r} +
V(r) - \varepsilon \right] u_{L}(r) \]

\n with boundary conditions such that its logarithmic derivative
matches that of the H\"ankel function. Any matrix element in this
basis then can be written as :

\begin{equation} \langle \phi_{iL} \vert {\cal{O}}
 \vert \phi_{jL'}\rangle \equal
\left[ \sum_{k} \int_{{\cal{S}}_{k}} \plus \int_{I} \right]
\phi_{iL}^{*}(\ul{r}){\cal{O}}\phi_{jL'}(\ul{r}) d^{3}\ul{r}
\end{equation}

The H\"ankel functions associated with a muffin-tin at
$\ul{R}_{i}$ can be written in terms of a Bessel function at
$\ul{R}_{j}$ as $\sum_{L^{\dpr}} {\cal{S}}_{iL,kL^{\dpr}}
J_{kL^{\dpr}}$ the structure matrix ${\cal{S}}$ depends
entirely on the geometric arrangement of the muffin-tins. The
first integral becomes :

 \begin{eqnarray} & = &
 \sum_{k\ne i,j}\sum_{L^{\dpr}}
{\cal{S}}_{iL,kL^{\dpr}}{\cal{S}}_{jL',kL^{\dpr}} \langle
J_{L^{\dpr}} \vert{\cal{O}} \vert J_{L^{\dpr}}\rangle_{S_{k}}  \plus
\sum_{k=i,\ne j}
{\cal{S}}_{jL',iL} \langle
H_{L} \vert {\cal{O}} \vert J_{L} \rangle_{S_{i}}
\ldots\nonumber\\
& & \plus \sum_{k=j,\ne i}
{\cal{S}}_{iL,jL'} \langle
J_{L} \vert {\cal{O}} \vert H_{L} \rangle_{S_{j}} \plus
\langle
H_{L} \vert {\cal{O}} \vert H_{L} \rangle_{S_{i}} \nonumber\\
& = & {\cal{O}}^{HH} \plus {\cal{S}}^{\dag} {\cal{O}}^{JH} \plus
{\cal{O}}^{HJ} {\cal{S}} \plus {\cal{S}}^{\dag}{ \cal{O}}^{JJ}
{\cal{S}}
\end{eqnarray}
These are easy to calculate and there is a {\sl separation} of
atomic and structural information.

Most of the interstitial integral can be obtained from the
muffin-tin spheres by using the fact that, in the interstitial,
the basis are solutions of the Helmholtz equation, and using the
Green theorem :

 \begin{eqnarray}
 \int_{I} \phi^{*}_{1} \phi_{2} d^{3}\ul{r} & \equal &
\frac{1}{\kappa_{1}^{2}-\kappa_{2}^{2}}\sum_{k}\int_{S_{k}}
\left[ \phi^{*}_{1}\nabla\phi_{2} -
\phi_{2}\nabla\phi_{1}^{*}\right] d^{2}\ul{r} \nonumber\\
 \int_{I} \phi^{*}_{1}\left(-\nabla^{2}\right) \phi_{2} d^{3}\ul{r}
 & \equal & \kappa_{2}^{2} \int_{I} \phi^{*}_{1} \phi_{2} d^{3}\ul{r}
 \end{eqnarray}

If the potential here is a constant we can get bye with the
above. But for clusters this is definitely not so. In the
molecular FPLMTO we use a tabulation technique. We expand the
product :

\[ \phi_{i}^{*}(\ul{r})\phi_{j}(\ul{r}) \equal \sum_{m}
C^{ij}_{m} \chi_{m}(\ul{m}) \]

\n where $\chi_{m}(\ul{r})$ is another set of muffin-tin centered
H\"ankel functions. In practice we put two atoms along the z-axis
and make accurate numerical expansion by least squares fit for
different distances and tabulate C$^{ij}_{m}$(d) :

\begin{eqnarray*}
A_{mn} & = & \int_{I} \chi_{m}^{*}(\ul{r})\chi_{n}(\ul{r})d^{3}r
\\
B_{m} & = & \int_{I}
\phi^{*}_{i}(\ul{r})\phi_{j}(\ul{r})\chi_{m}(\ul{r}) d^{3}\ul{r}
\\
{\cal{C}} & = & {\cal{A}}^{-1}{ \cal{B}}
\end{eqnarray*}

\n This is the two-centre fit table (TCF). For arbitrary geometry then
we may easily calculate the necessary matrix elements by a
fitting procedure to the table. The procedure id {\sl fast}.

For molecular dynamics, the problem arises from the fact that the
Pulay terms in the force are impossibly difficult to calculate
directly as the basis set changes in a complicated manner when
atoms move. To do the molecular dynamics, we use the Harris
functional procedure as follows : At a time step $\tau_{0}$ we
obtain the self-consistent charge density $\rho(r,\tau_{0})$
using the FP-LMTO procedure. At a neighbouring time
$\tau_{0}$+$\tau$ we hazard a guess $\rho_{g}(r,\tau_{0}+\tau)$
and obtain
\begin{eqnarray*}
\~E(\tau)& = & E_{H}[\rho_{g}(r,\tau_{0}+\tau)]\\
& = &
\sum_{i}\varepsilon_{i}\left[V_{eff}\right] - \int
\rho_{g}(\ul{r}) V_{eff}\left[\rho_{g}(\ul{r})\right] + U\left[
\rho_{g}(\ul{r})\right] + E_{xc}\left[\rho_{g}(\ul{r})\right] \end{eqnarray*}

To find the force on an atom, we simply move the atom with its surrounding
charge density in a given direction. 
The force is given by $$\frac{\partial\~E}{\partial\tau}
\rule[-5mm]{0.2mm}{11mm}\; _{\tau\rightarrow 0}$$

For {\sl dynamics} we use the Verlet algorithm :

\[ \ul{r}_{n+1} \equal 2\ul{r}_{n}-\ul{r}_{n-1} +
\frac{\ul{F}}{m} (\Delta t)^{2} \]

\n where n denotes the time step of length $\Delta t$. We can now
either do straightforward molecular dynamics, but this often
leads to unphysical heating/cooling of the system if our time
steps are too large. For small time steps the procedure is
inordinately slow. We add an extra
friction term carefully $\ul{F} \Rightarrow  \ul{F} - \gamma
m\ul{\dot{r}}$. Methfessel and Schilfgaarde \cite{kn:ms} have also used a free
dynamics with feedback to overcome the above difficulty.

\section{Results}

We have chosen the various parameters for the FP-LMTO based on
optimizing results for  bulk Cu and the dimer. The values of
$\kappa^{2}$ were chosen from optimum bulk calculations. The
muffin tin radii were chosen as 1.9 \AA to produce the bond
length and binding energies of the Cu dimer correctly. For
augmentation within the sphere we have used 4$s$, 4$p$, 3$d$, 4$f$ and 5$g$
functions ($\ell_{max}$=3). For representation of interstitial
functions we have used five $\kappa^{2}$ values with angular
momentum cutoffs $\ell_{max}$ = 4,4 6,2 and 1.

\begin{table}
\centering
\caption{Bond lengths and Binding energies/atom for {\sl Cu$_2$} dimer}
\label{tab2}
\vskip 0.5cm
\begin{tabular}{c|c|c|l}\hline\hline
Bond length& Binding Energy &  Method & Reference \\
  (a.u.)   &   (eV/atom)    & & \\ \hline\hline
\multicolumn{4}{c}{Theoretical Results}\\ \hline
4.55 & - & ab initio SCF  & \cite{miyo} \\
4.28 & 0.923   & LSD Pseudo.      & \cite{wang} \\
4.20 & 1.025   & LSD Pseudo.\ +\ SIC   & \cite{wang} \\
4.28 & 0.975   & LSD Pseudo.\ +\ CI & \cite{wang}\\
4.56 & 0.34    & Hartree-Fock & \cite{amos}\\
4.55 & 1.04    & CI           & \cite{amos}\\
4.20 & 1.50    & $X-\alpha$   & \cite{amos}\\
4.17 & 1.30    & LCGO-DFT     & \cite{ckrs} \\
4.30 & 1.13    & LCGO-GGA     & \cite{ckrs}\\ 
4.16 & 1.369   & FP-LMTO-TCF  & Our work \\ 
  -  & 0.23    & TB-LMTO      & \cite{kn:lmto}\\ \hline
\multicolumn{4}{c}{Experimental Results}\\ \hline
4.195 &  0.99  &  Expt & \cite{kn:huber} \\
4.21 & 1.03 & - & \cite{abr}\\
  -  & 1.04 &   -  & \cite{rv} \\ \hline\hline
\end{tabular}
\end{table}

The optimum bond length was determined by varying the dimer bond length
from 4.1 to 4.2 atomic units and calculating the total energy at each bond length.
We found the optimum bond length to be 4.16 a.u with the binding energy (B.E) 
equal to 1.469 eV/atom. The table \ref{tab2} lists the various theoretical
and experimental values for the bond length and binding energies per atom. It
is well known that while the Hartree-Fock tends to under-bind, the LDA over-binds.
Our bond lengths should then be smaller and binding energies larger than experimental
values. This is borne out by the table. Clearly both the self interaction correction
(SIC) and the gradient correction (GGA) improves matters. 
The TB-LMTO value of 0.23 eV/atom (\cite{kn:lmto}) is much too low and probably
indicates serious lacun\ae\ in the treatment of clusters in that work rather than
in the TB-LMTO itself.

\begin{table}
\centering
\caption{Ionization potentials for { Cu$_2$} by various methods}
\label{tab1}
\vskip 0.4cm
\begin{tabular}{c|c|l}\hline\hline
IP (eV) & Method & Reference \\ \hline\hline
5.65    & ab initio SCF $\Delta E$ & \cite{miyo} \\
6.04    & ab initio SCF Koopman    & \cite{miyo} \\
7.987   & LSD Pseudopotential       & \cite{wang} \\
8.237   & LSD Pseudopotential\ +\ SIC   & \cite{wang} \\
6.37    & Hartree-Fock              & \cite{amos} \\
7.37    & Modified Hartree-Fock-I   & \cite{amos} \\
7.89    & Modified Hartree-Fock-II  & \cite{amos} \\
7.35    & $X-\alpha$                & \cite{amos} \\
7.64    & EHT                       & \cite{amos} \\
5.70    & LCAO-SCF                  & \cite{cv} \\
8.69    & LCGO-DFT                  & \cite{ckrs}\\
7.904   & LCGO-DFT\ +\ GGA              & \cite{ckrs} \\
8.22    & FP-LMTO-TCF               & Our work \\ \hline\hline
\end{tabular}
\end{table}

The table \ref{tab1} shows the ionization potential (IP) for $Cu_2$ dimers, calculated
as the difference between the  total energies of neutral $Cu_2$ and the $Cu_{2}^{+}$ ion,
using various methods. The experimental values quoted range between 7.904$\pm$0.04 
quoted by Calamici \etal \cite{ckrs} and 7.37 of Joyes and Leleyer \cite{jl}.
It is quite clear that for the smaller
clusters the generalized gradient corrections (GGA) to the local density approximation is
very important (\cite{ckrs}). Our FP-LMTO does not incorporate the GGA and hence leads
to slightly larger values of the IP. The importance of self-interaction corrections (SIC)
is not clear for dimers. Wang \cite{wang} includes SIC and obtains a higher value of the IP.
Our work does not include the SIC.

The first test of the predictability of various methods first appear for $Cu_3$. The
accompanying figure 1 shows the lowest energy structures predicted for the trimer. 
Miyoshi \etal \cite{miyo} find both the structures (O) and (A) to be almost degenerate in energy. The vertex
angles are found to be 77.6$^{o}$ for (O) and 51.7$^{o}$ for (A). Calamici \etal \cite{ckrs} find the structure
(O) to be most stable with vertex angle 66.86$^{o}$ without SIC and 66.58$^{o}$ with SIC. The
other structure (A) lies 0.023 eV higher in energy. Wang \cite{wang} finds the obtuse triangle
 shown on the right to be the stable structure. This has a vertex angle of 162$^{o}$. He
concludes that the SIC correction is essential and finds the acute triangle with a vertex
angle of 47$^{o}$ to be the most stable. However, even with the SIC the structure quoted
is rather different from other methods. Our prediction agrees reasonably well with the
structure (O) of Miyoshi \etal \cite{miyo} and (O) of Calamici \etal
\cite{ckrs}. The vertex angle is 65$^{o}$ in our
case. The isosceles shape is expected because of the possible Jahn-Teller distortion in
$Cu_3$.

\begin{table}
\centering
\caption{Bond lengths, binding energies and ionization potentials of { Cu$_3$}}
\label{tab3}
\vskip 0.4cm
\begin{tabular}{c|c|c|l}\hline\hline
Side length& Binding Energy & IP &   Reference \\
  (a.u.)   &   (eV/atom)    & (eV)  & \\ \hline\hline
\multicolumn{4}{c}{Theoretical Results}\\ \hline
4.72 & 1.018 & 4.39 &  \cite{miyo} \\
4.50 & 0.942 & 7.144 &  \cite{wang} \\
5.41 & 0.753 & 6.018 &  \cite{wang} \\
4.35 &  -    & 6.33 &  \cite{amos} \\
4.28 & 1.34 & 6.46 &  \cite{ckrs}\\
4.45 & 1.12 & 5.795 &  \cite{ckrs}\\
  -  & 0.68 &  -    &  \cite{kn:lmto} \\
4.30 & 1.598 & 6.40 &  Our work \\ \hline 
\multicolumn{4}{c}{Experimental Results}\\ \hline
 -    & 1.02  & 5.48$\pm$0.5 &  \cite{wvz} \\
\hline\hline \end{tabular}
\end{table}

The table \ref{tab3} compares the bond lengths, binding energies and ionization
energies of he $Cu_3$ trimer. We find 
the binding energy per atom to be 1.598 eV/atom which is higher than that for the linear 
configuration by 0.124 eV/atom.  Over-binding because of the LDA is again observed.
The ionization energy drops for the trimer and regains its value again for $Cu_4$. This
has been observed in all the earlier works quoted and in experiment.

For N=4 we find the rhombus starting structure to lead to the most stable 
structure followed by the square and the tetrahedron in decreasing order
of stability. 
 Our prediction matches exactly with that of Akeby \etal \cite{kn:akeby} and  Calamici \etal 
\cite{ckrs}  who also predicted
the sequence rhombus, square and tetrahedron. The larger rhombus angle turns out
to be 120$^{o}$ which agrees well with the prediction of 122$^{o}$ by Calamici \etal \cite{ckrs}. Our
ionization potential is 7.90 eV, which agrees not badly with 7.0$\pm$0.6 eV found
experimentally.
 The TB-LMTO (\cite{kn:lmto}) predicts the order of stability to be the tetrahedron,
the rhombus and the square  in decreasing stability. This does not match with any
other work and possibly has its origin in the problem talked about earlier.

For $Cu_{5}$ we find the trigonal bipyramid with B.E. 2.187 eV/atom to be the most 
stable structure followed by the square pyramid where the difference in B.E. 
between the two structures is .056 eV only. 
 Akeby \etal \cite{kn:akeby} also 
obtain the trigonal bipyramid to be more stable than the square pyramid
agreeing with our calculations. Calamici \etal \cite{ckrs} finds another structure, the
flat pentagonal trapezoid to be almost degenerate; actually 0.009 eV lower
in energy than the trigonal bipyramid. They find the square pyramid to be
more than 0.309 eV higher in energy. We would like to emphasize with Calamici \etal
that for the smaller clusters the GGA may play a crucial role in stabilizing
certain structures.

Figure 3 shows the variation of the ionization potential with cluster size. The
troughs at n=3 and n=5 agree well with earlier works as well as experimental results
(\cite{knik}).

For N=6 we have considered two  starting structures the square bipyramid
(octahedron) and the capped trigonal bipyramid which is obtained by
capping one face of the trigonal bipyramid so that the capping atom is 
equidistant from all the three atoms on the face. We find the capped 
trigonal bipyramid to be the most stable structure with bond energy equal
to 2.405 eV/atom which is 0.040 eV/atom higher than the square bipyramid
(octahedron). The TB-LMTO calculations predict the octahedron to be
the most stable structure compared to other random structures. Also
the numerical value of 1.56 eV/atom for the octahedron obtained
from the TB-LMTO calculations is much lower compared to our value.
         
The pentagonal bipyramid, the capped square bipyramid and the bicapped
trigonal bipyramid were considered as the starting structures for our 
calculations for N=7. We find the pentagonal bipyramid to be the most 
stable structure in accordance with Akeby \etal.
but at variance with the TB-LMTO results. The bicapped trigonal
bipyramid is slightly higher in energy (0.002 eV/atom) than the capped
square pyramid in our calculations.

For $Cu_{8}$ we considered three starting structures as shown in the
table of which the capped pentagonal bipyramid turns out to be the most stable
followed by the bicapped square bipyramid and the cube. TB-LMTO predicts the 
antiprism followed by the bi-tent structure and the cube. Both the methods
find the cube to b ethe least stable though our B.E. for the cube is
0.592 eV/atom higher than the TB-LMTO results.

In the case of $Cu_9$,  we considered the tricapped square bipyramid
and the bicapped pentagonal bipyramid with the capping atoms on adjacent
and non-adjacent faces. The tricapped square bipyramid was found to be the most
stable structure followed by the bicapped pentagonal bipyramid with
the capping atoms on adjacent faces (lower by only 0.006 eV/atom) and the
bicapped pentagonal bipyramid (non-adjacent faces) lower by 0.025 eV/atom
than the most stable structure in this range.
The stable shapes for 6 $\leq$ N $\leq$ 9 are shown in figures 4.

Figure 5 shows a plot of the binding energy versus cluster size for N=2 to 9.
The relative stability of the clusters ($2E(N)-E(N+1)-E(N-1)$)
is also plotted on the same graph. Cluster sizes N=3,5 and 8
show up as more stable. This has been predicted experimentally earlier by Knickelbein \cite{knik}.
Katakuse \cite{kn:kat} has also observed N=8 to be a stable structure
in their experimental observations.

Figure 6  shows a plot for the HOMO-LUMO gap versus cluster size
for the most stable clusters. In the theoretical results of Akeby \etal 
 \cite{kn:akeby} the HOMO-LUMO gap for the $Cu_{8}$ cluster is
determined to be 1.93 eV while  Lammers and Borstal \cite{kn:lmto} report
a value of 1.91 eV. Our calculated value for the HOMO-LUMO gap for the 
most stable structure (capped pentagonal bipyramid) is 1.156 which
is lower than both the reported values. The HOMO-LUMO gap does show a peak
at N=8 in our calculations but we cannot conclude from this point that 
this is a manifestation of shell closure. We also see a minimum in the 
HOMO-LUMO gap value at N=6 unlike in \cite{kn:lmto}. Moreover pronounced
odd-even alterations in the HOMO-LUMO gap values as predicted by the shell
model (\cite{kn:sugano}) are not recognizable in our calculations.

The N=10 cluster shape is a close competition between the tetracapped trigonal bipyramid
which is obtained by capping the N=9 cluster on another face and the structure shown on
the right hand side of figure 6. Our calculations indicate that the former
is more stable, however, the energy difference is smaller than the errors involved in
the FP-LMTO itself. From N=11 to N=13 the clusters grow towards the stable icosahedron.
These shapes indicate that probably our prediction is valid.

 For $N=12$ we started from a configuration which is an  
icosahedron with a void at the centre. Rapidly the structure
evolved to the icosahedron with one exterior atom removed. 

For $N=13$ we studied carefully two possible structures : the cubo-octahedron (shown
on the left in figure 9 and the icosahedron, shown on the right of the
same figure. Our calculations indicate that even if we begin with the cubo-octahedron
as our starting structures, the cluster rapidly settles down to the icosahedron. Earlier
Valkealahti and Manninen \cite{v92} had also used effective medium-molecular 
dynamics and shown that the cubo-octahedron
is unstable and rapidly changes over to the stable icosahedron. Winter \etal \cite{wpr} have argued
from experimental observations that the shell structure seen in the smaller clusters is
overshadowed by icosahedral closures from $N=13$ onwards.

For $N=15$ and $N=16$ we see near-degenerate structures. The
lower-energy structure has atoms on neighboring faces of the
icosahedron. There is also another structure, differing in
energy by about $1\%$, in which the ``extra" atoms are on
non-neighboring faces of the icosahedron. For $N=17$ the two
different starting structures both anneal to an icosahedron
with four atoms on neighboring faces.

The $N=19$ has a very stable structure : the double icosahedron, further confirming
the conjecture of Winter \etal \cite{wpr} regarding icosahedral closure.
 For $N=20$ The equatorial addition was
found to be more stable by about $1\%$. We expect as the size increases,
the cluster structure becomes more spherical. Note that we see no
evidence for the very open structure reported to have been
obtained by Lammers and Borstal \cite{kn:lmto} for $N=20$ through simulated annealing.

Figure 12 shows the binding energy and the homo-lumo gap for the clusters
$N=11$ to $N=20$. We note that the signatures of shell closure we observed in the smaller
clusters is overtaken by geometric closures and the icosahedron based closed structures
are the more stable.

\section*{\bf Acknowledgements}

We should like to thank Profs. M. Methfessel and M. Van
Schilfgaarde for making the entire mechanism of the FP-LMTO
available to us and enthusing us to make use of this powerful
technique.

\newpage

\centerline{\Large\bf FIGURE CAPTIONS}
\vskip 1cm
\begin{description}
\item[1]{Various shape predictions for the {\sf Cu$_3$} trimer}
\item[2]{Stable configurations for {\sf Cu$_4$} and {\sf Cu$_5$}}
\item[3]{Variation of the ionization potential with cluster size}
\item[4]{Stable configurations for {\sf Cu$_6$} to {\sf Cu$_9$}}
\item[5]{The binding energy per atom and its curvature for {\sf Cu$_2$} to {\sf Cu$_9$}}
\item[6]{The homo-lumo gap for {\sf Cu$_2$} to {\sf Cu$_9$}}
\item[7]{The structures for {\sf Cu$_{10}$}}
\item[8]{The stable structures for {\sf Cu$_{11}$} and {\sf Cu$_{12}$}}
\item[9]{The structures for {\sf Cu$_{13}$ }}
\item[10]{The structures for {\sf Cu$_{14}$} -- {\sf Cu$_{18}$}}
\item[11]{The structures for {\sf Cu$_{19}$} -- {\sf Cu$_{20}$}}
\item[12]{The binding energy per atom and the homo-lumo gap for {\sf Cu$_{11}$} to {\sf Cu$_{20}$}}
\end{description}
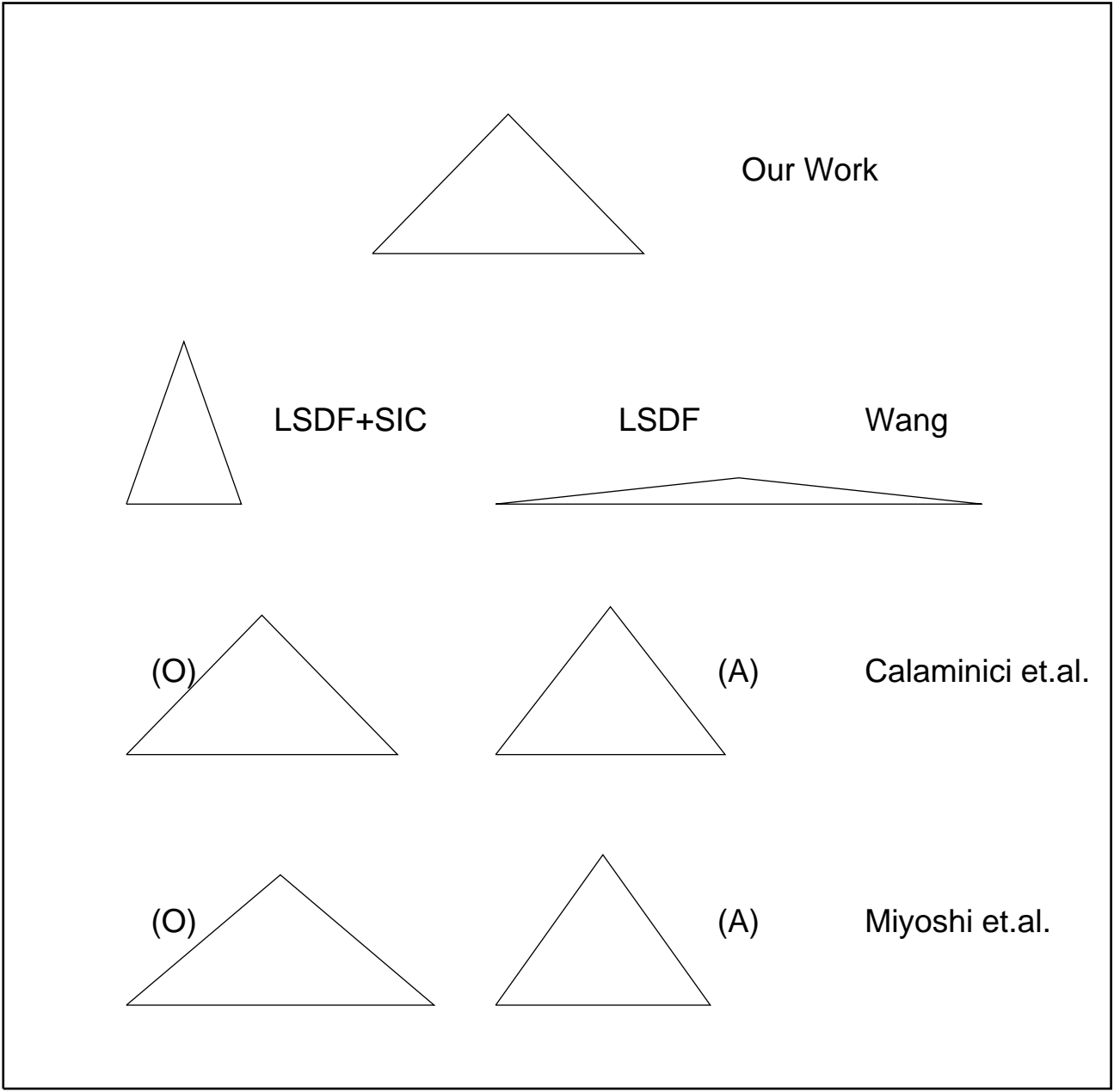
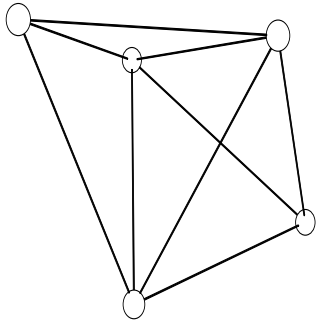
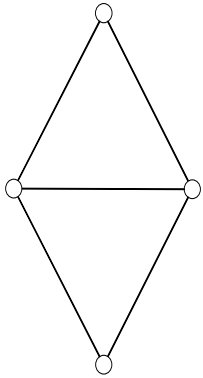
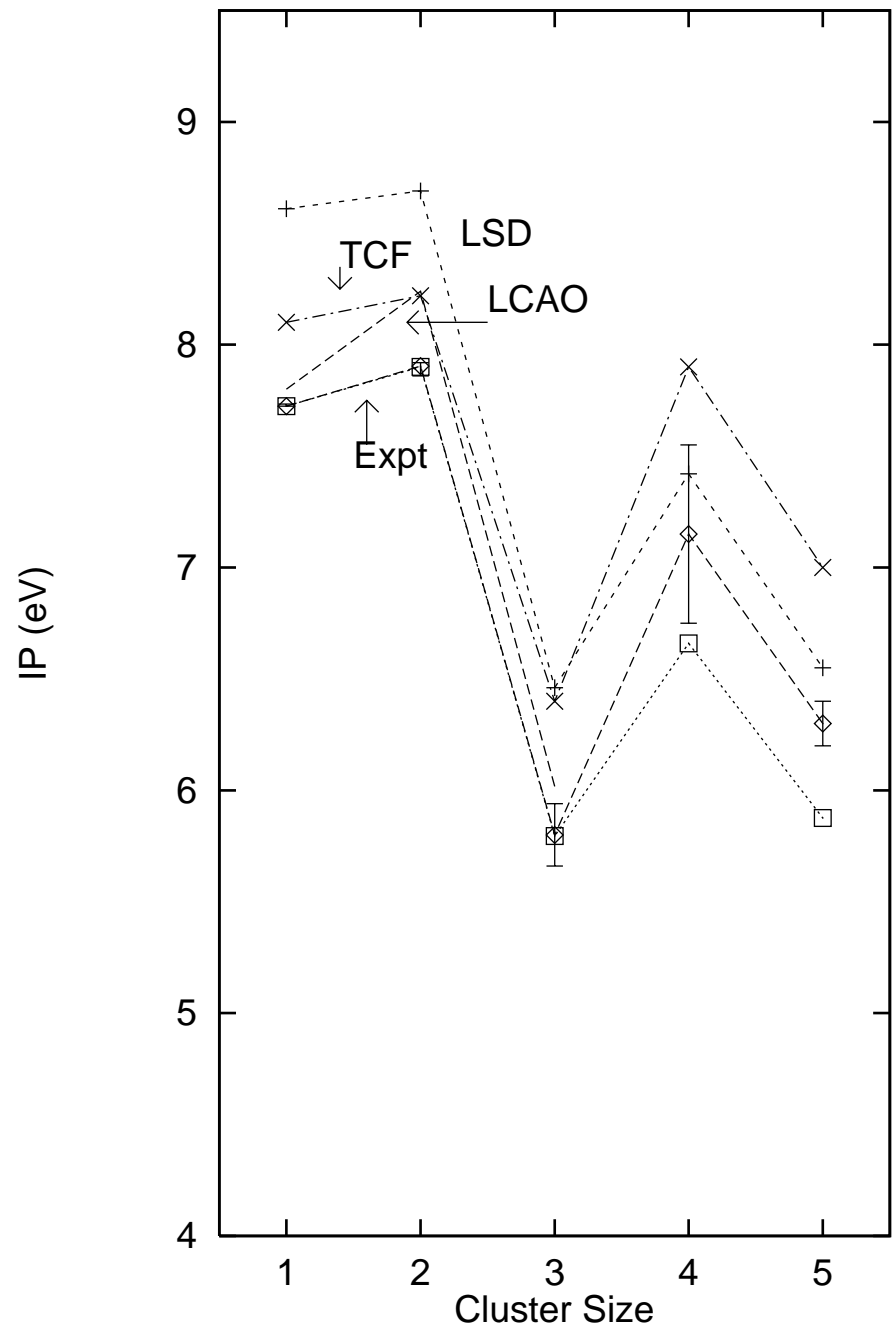
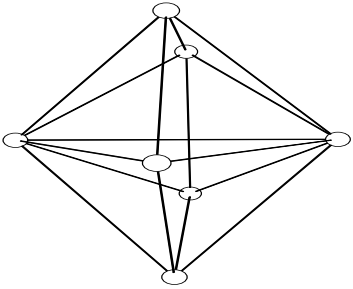
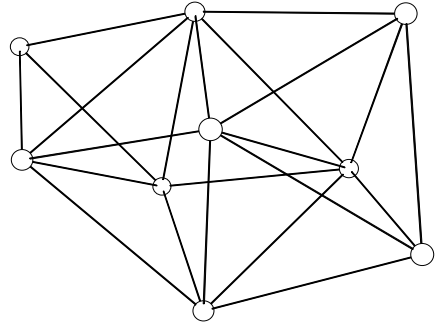
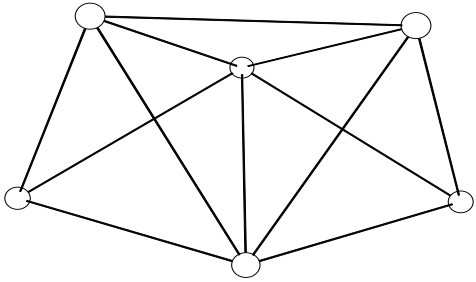
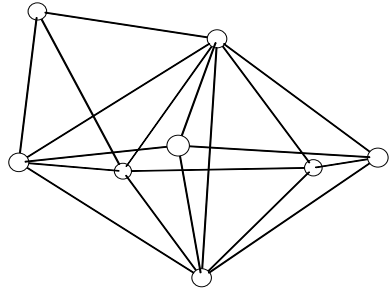
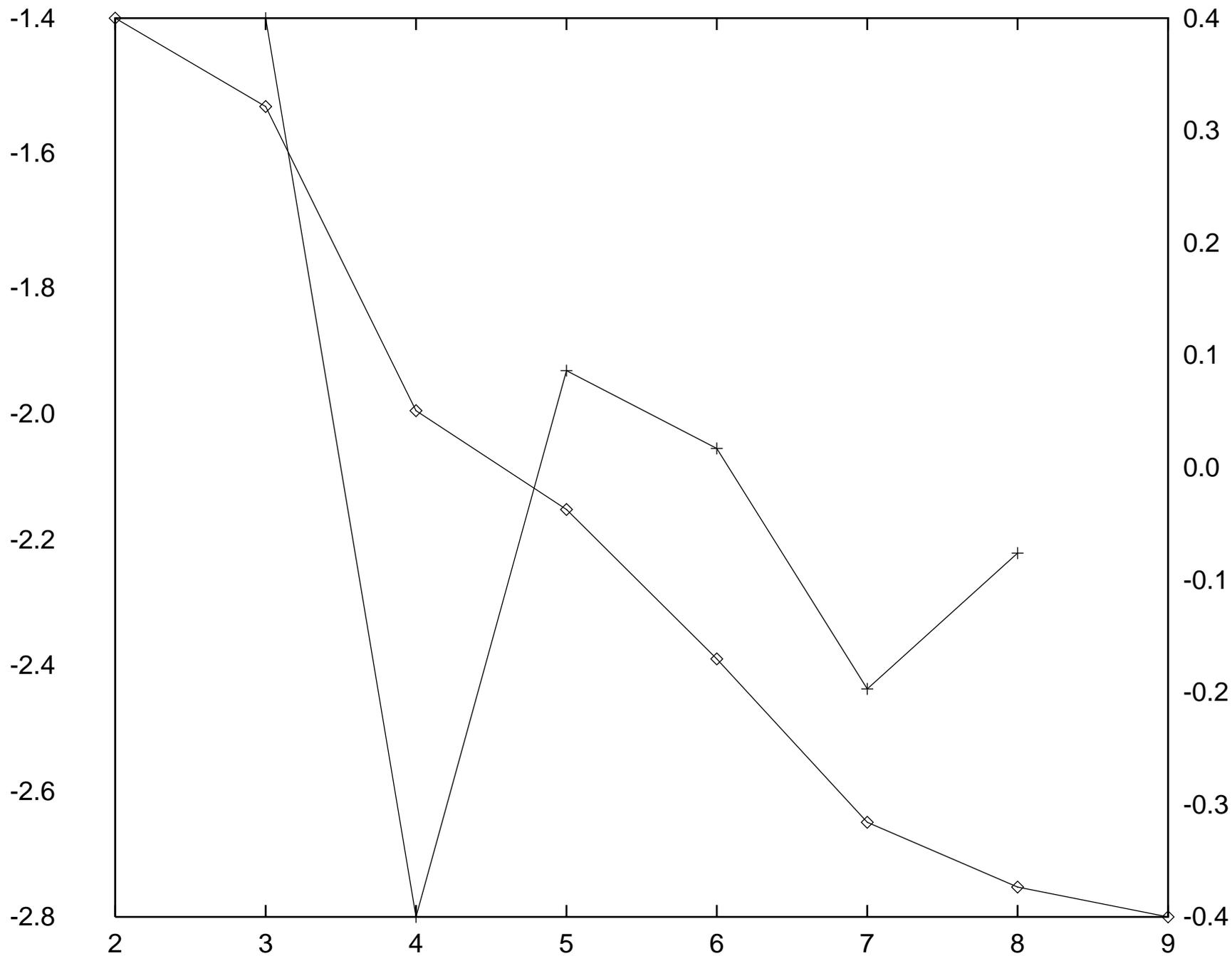
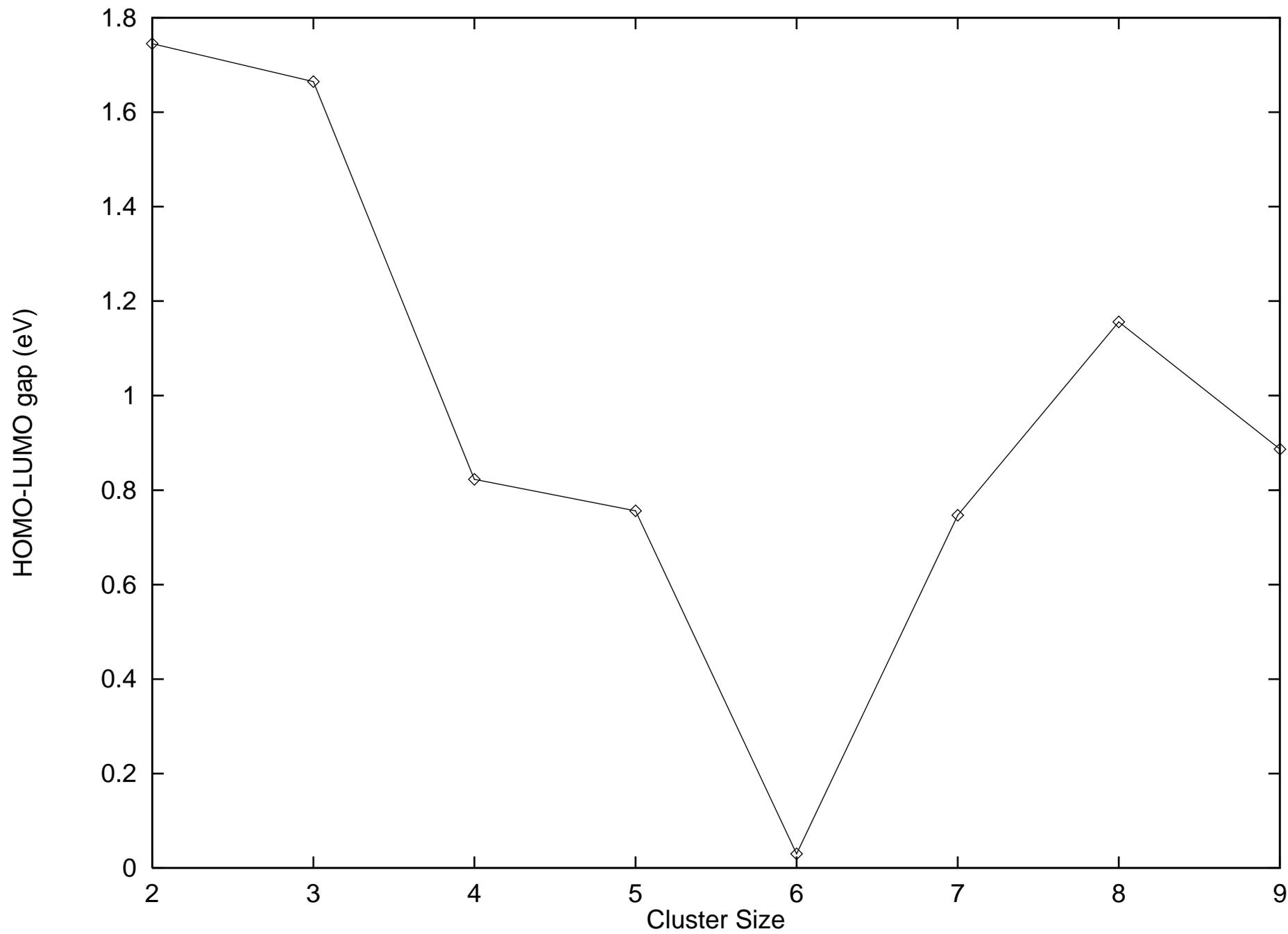
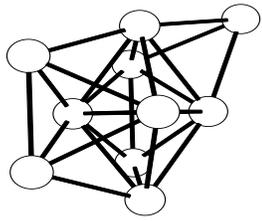
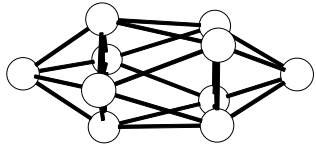
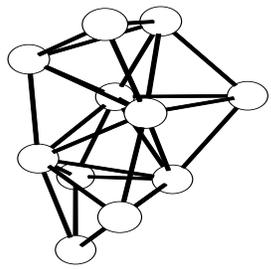
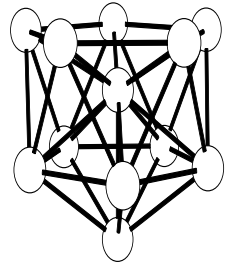
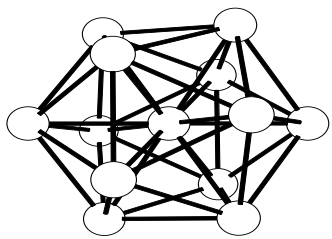
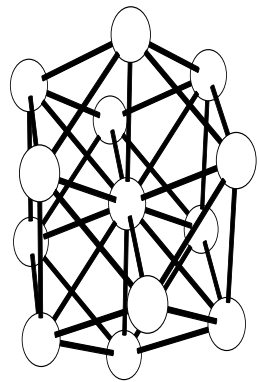
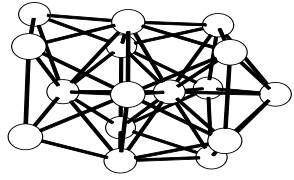
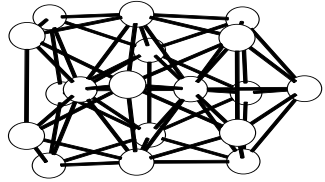
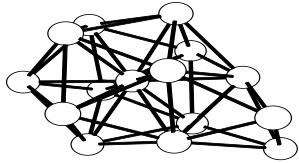
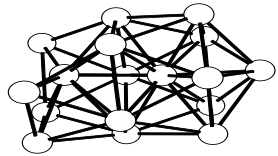
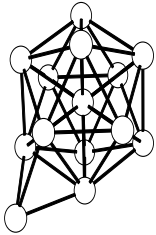
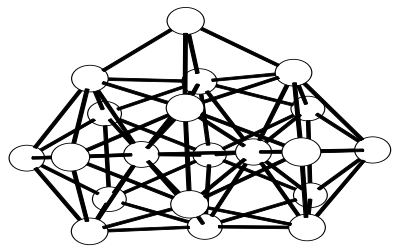
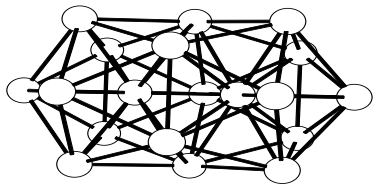
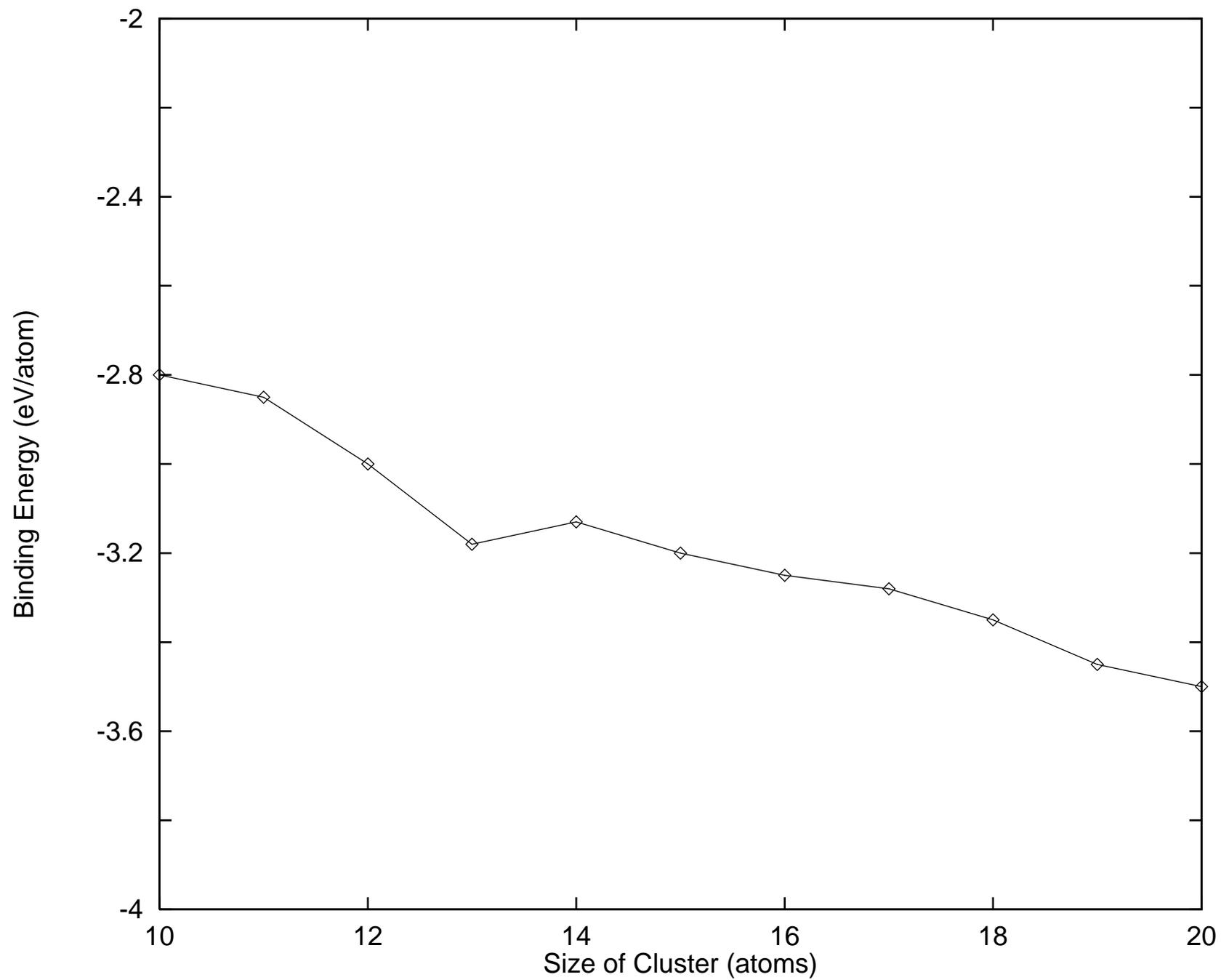
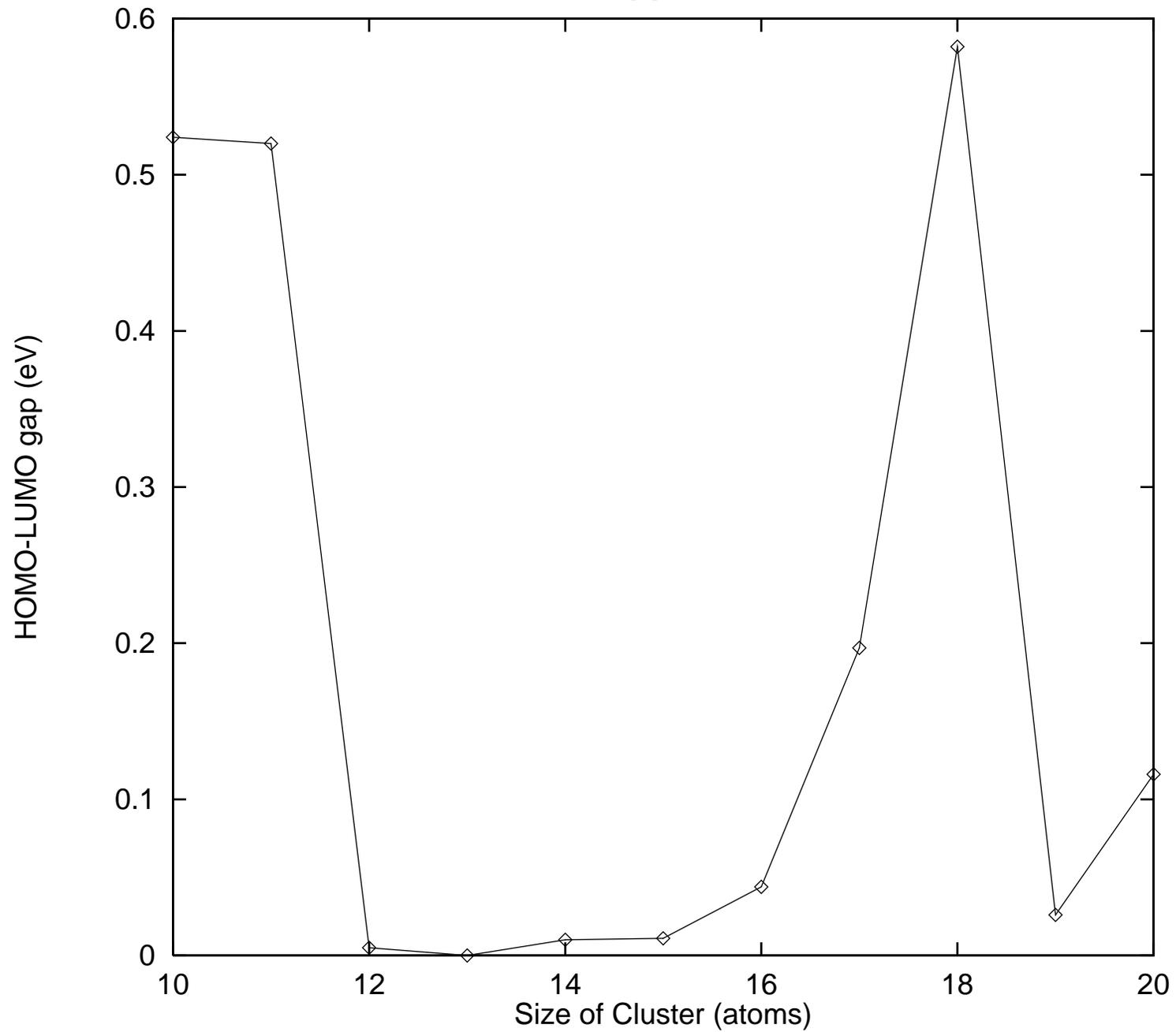
\end{document}